\documentclass[prl,aps,twocolumn,superscriptaddress,%showpacs
]{revtex4}
\usepackage{amsmath,amssymb,graphicx}
\usepackage{pxfonts}
\usepackage{graphicx}
\usepackage{color}
\usepackage{float}
\begin{document}

\date{\today}

\title{Interference dislocations adjacent to emission spot}

\author{J.~R.~Leonard}
\affiliation{Department of Physics, University of California San Diego, La Jolla, CA 92093, USA}
\author{L.~H.~Fowler-Gerace}
\affiliation{Department of Physics, University of California San Diego, La Jolla, CA 92093, USA}
\author{Zhiwen~Zhou}
\affiliation{Department of Physics, University of California San Diego, La Jolla, CA 92093, USA}
\author{E.~A.~Szwed}
\affiliation{Department of Physics, University of California San Diego, La Jolla, CA 92093, USA}
\author{D.~J.~Choksy}
\affiliation{Department of Physics, University of California San Diego, La Jolla, CA 92093, USA}
\author{L.~V.~Butov} 
\affiliation{Department of Physics, University of California San Diego, La Jolla, CA 92093, USA}

\begin{abstract}
\noindent We studied interference dislocations (forks) adjacent to an emission spot in an interference pattern. The adjacent interference dislocations are observed in emission of excitons in a monolayer transition metal dichalcogenide and in emission of spatially indirect excitons, also known as interlayer excitons, in a van der Waals heterostructure. The simulations show that the adjacent interference dislocations appear due to the moir{\'e} effect in combined interference patterns produced by constituting parts of the emission spot. The adjacent interference dislocations can appear in interference images for various spatially modulated emission patterns. 
\end{abstract}
\maketitle

Dislocations in interference patterns are studied in a variety of systems including photons~\cite{Scheuer1999}, 
atoms~\cite{Inouye2001, Hadzibabic2006}, 
polaritons~\cite{Lagoudakis2008, Lagoudakis2009, Sanvitto2010, Roumpos2011, Amo2011, Grosso2011, Hivet2012, Flayac2013, Bardyn2015}, 
magnons~\cite{Nowik-Boltyk2012}, 
and excitons~\cite{High2012, Vishnevsky2013, Leonard2018, Leonard2021}. 
Interference dislocations can originate from quantized vortices and vortex-related interference dislocations were reported for optical vortices~\cite{Scheuer1999}, atom vortices~\cite{Inouye2001, Hadzibabic2006}, polariton vortices~\cite{Lagoudakis2008, Lagoudakis2009, Sanvitto2010, Roumpos2011}, and magnon vortices~\cite{Nowik-Boltyk2012}. Interference dislocations can also originate from the moir{\'e} effect in interference patterns~\cite{Leonard2021}. In the case of long-range ballistic propagation of matter waves to the locations of interference dislocations, the moir{\'e}-effect interference dislocations can evidence an anomalously long mean free time and, in turn, superfluidity in the considered system~\cite{Leonard2021}. These isolated interference dislocations originating from the moir{\'e} effect in combined interference patterns of propagating condensate matter waves were observed for condensate of indirect excitons (IXs) long distances away from the IX sources~\cite{Leonard2021}.

\vskip5mm
{\bf Experiments}

In this work, we study a different type of interference dislocations originating from the moir{\'e} effect -- the interference dislocations adjacent to the exciton source and emission spot. To consider these emission-spot-adjacent interference dislocations, further called adjacent interference dislocations for simplicity, we measured interference patterns for excitons in a WSe$_2$ monolayer and for IXs in a MoSe$_2$/WSe$_2$ van der Waals heterostructure. IXs are composed from electrons and holes in separated layers~\cite{Lozovik1976} and IXs in the MoSe$_2$/WSe$_2$ heterostructure are formed by electrons in MoSe$_2$ monolayer and holes in WSe$_2$ monolayer~\cite{Rivera2015}. The heterostructure details are presented in Supplementary Information (SI)~\cite{SI}.

Figure~1 shows the interference pattern of exciton emission measured by shift-interferometry. The measurements are outlined below. Excitons transform to photons as a result of exciton recombination. The photons go through the Mach-Zehnder interferometer. Each of the two interferometer arms forms an image of the exciton emission on CCD detector. These two images are shifted relative to each other in the $x$-direction and the interference pattern is produced by the interference between the emission of excitons separated by the shift $\delta x$ in the heterostructure plane. The measurement details are presented in SI.

Figure~1 shows interference dislocations in the interference pattern of excitons in a WSe$_2$ monolayer. These interference dislocations are observed close to the exciton emission spot and are adjacent interference dislocations. Figure~2 shows adjacent interference dislocations in the interference pattern for IXs in a MoSe$_2$/WSe$_2$ heterostructure.

\begin{figure}
\begin{center}
\includegraphics[width=8.5cm]{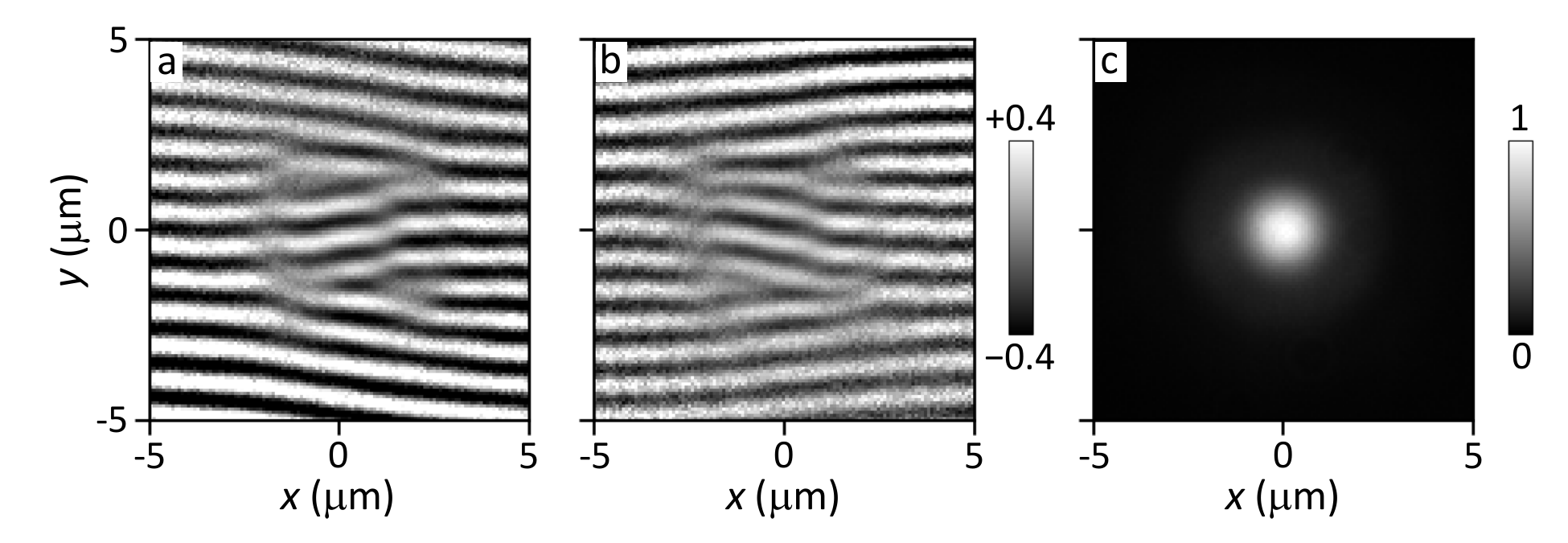}
\caption{Adjacent interference dislocations in interference pattern for excitons in WSe$_2$ monolayer. 
(a, b) Measured interference pattern $I(x, y)$ for the shift $\delta x = - 1.4$~$\mu$m (a) and $\delta x = 1.4$~$\mu$m (b). 
(c) The exciton emission spot. 
The adjacent interference dislocations are observed close to the exciton emission spot.
}
\end{center}
\label{fig:spectra}
\end{figure}

\begin{figure}
\begin{center}
\includegraphics[width=8.5cm]{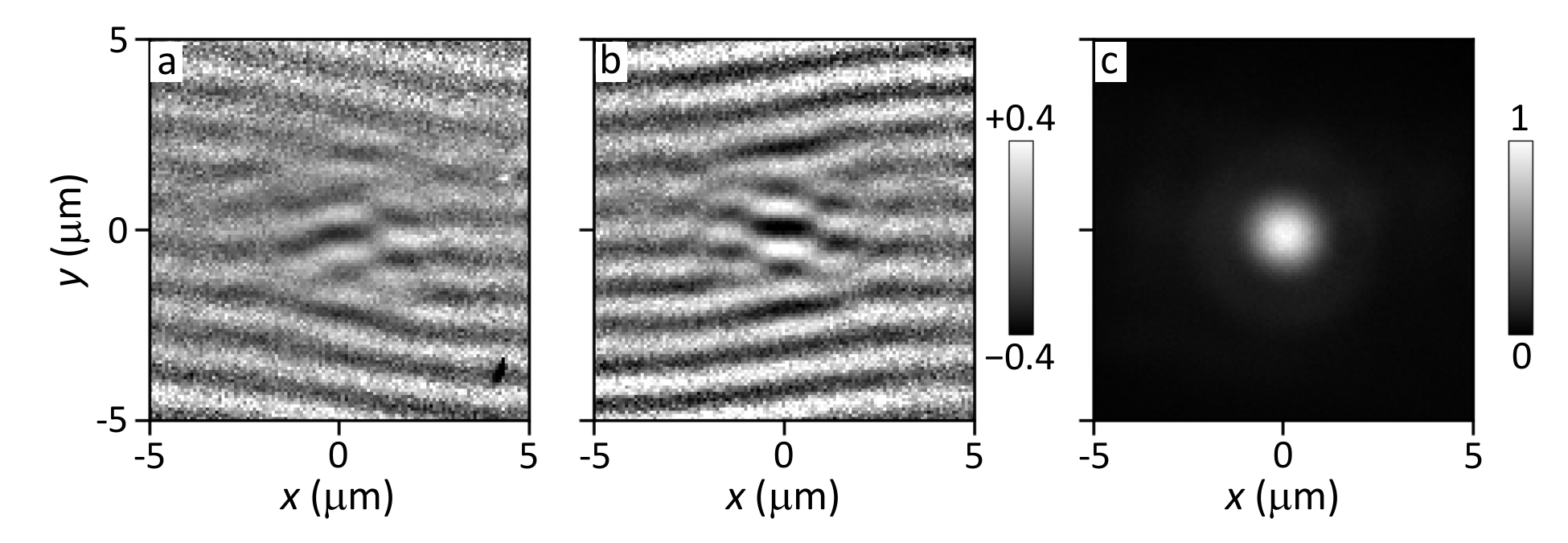}
\caption{Adjacent interference dislocations in interference pattern for indirect excitons (IXs) in MoSe$_2$/WSe$_2$ heterostructure. 
(a, b) Measured interference pattern $I(x, y)$ for the shift $\delta x = - 1.4$~$\mu$m (a) and $\delta x = 1.4$~$\mu$m (b). 
(c) The IX emission spot. 
The adjacent interference dislocations are observed close to the IX emission spot.
}
\end{center}
\label{fig:spectra}
\end{figure}

\vskip5mm
{\bf Simulations and discussions}

First, we clarify a relation between the observed interference dislocations and vortices. For vortex-related interference dislocations, the distance between left- and right-dislocations is given by the shift in the shift-interferometry experiment~\cite{Leonard2021}. However, the distance between left- and right-dislocations in Figs.~1 and 2 is significantly different from the shift showing that the observed interference dislocations are not associated with vortices.  

Our simulations producing adjacent interference dislocations are outlined below. We consider the exciton emission spot as a superposition of emissions of point sources with the intensities $P(\mathbf{r}_{\rm s})$ given by the spatial profile of the excitation spot where excitons are generated. $\mathbf{r}_{\rm s}$ is the location of the point source in the excitation spot. 
We approximate each of the point source of excitons by a radial wave with $\psi({\bf r}) \propto e^{i k R}$, where $R=|\mathbf{r} -\mathbf{r}_{\rm s}|$ is the distance to the point source and $\mathbf{k}=k (\mathbf{r}-\mathbf{r}_{\rm s})/R$ is the exciton momentum. We consider the combined emission pattern produced by the superposition of point sources constituting the excitation spot $I_{\rm em}(\mathbf{r}) \sim \sum_{\mathbf{r}_{\rm s}} P(\mathbf{r}_{\rm s})/|\mathbf{r} -\mathbf{r}_{\rm s}|$. 

For the shift-interferometry experiments with the shift $\delta {\bf r}$, a point source of excitons with wave function $\psi({\bf r})$ produces the interference pattern $I_{\rm interf}({\bf r}) = \vert \psi({\bf r} - \delta {\bf r} / 2) e^{iq_t y} + \psi({\bf r}+\delta {\bf r} / 2) \vert^2$, where $q_t = 2\pi\alpha/\lambda$ gives the period of the interference fringes, $\alpha$ is a small tilt angle between the image planes of the interferometer arms, and $\lambda$ is the emission wavelength. We consider the combined interference pattern produced by the superposition of point sources constituting the excitation spot

\begin{eqnarray}
I_{\rm interf}(\mathbf{r}) \sim \sum_{\mathbf{r}_{\rm s}} \frac {P(\mathbf{r}_{\rm s})}{|\mathbf{r} -\mathbf{r}_{\rm s}|} \vert \psi({\bf r} - \delta {\bf r} / 2) e^{iq_t y} + \psi({\bf r}+\delta {\bf r} / 2) \vert^2.
\end{eqnarray}

For the shift $\delta {\bf r} = \delta {\bf x}$, the interference pattern produced by each point source generating $\psi({\bf r}) \propto e^{i k R}$ is given by $I_{\rm interf}(\mathbf{r}, \mathbf{r}_{\rm s}) \sim \cos({\bf k} \delta {\bf x} +q_ty)$ and the interference pattern described by Eq.~1 is given by

\begin{eqnarray}
I_{\rm interf}(\mathbf{r}) \sim \sum_{\mathbf{r}_{\rm s}} \frac {P(\mathbf{r}_{\rm s})}{|\mathbf{r} -\mathbf{r}_{\rm s}|} \cos ({\bf k} \delta {\bf x} + q_t y).
\end{eqnarray}

\begin{figure}
\begin{center}
\includegraphics[width=8.5cm]{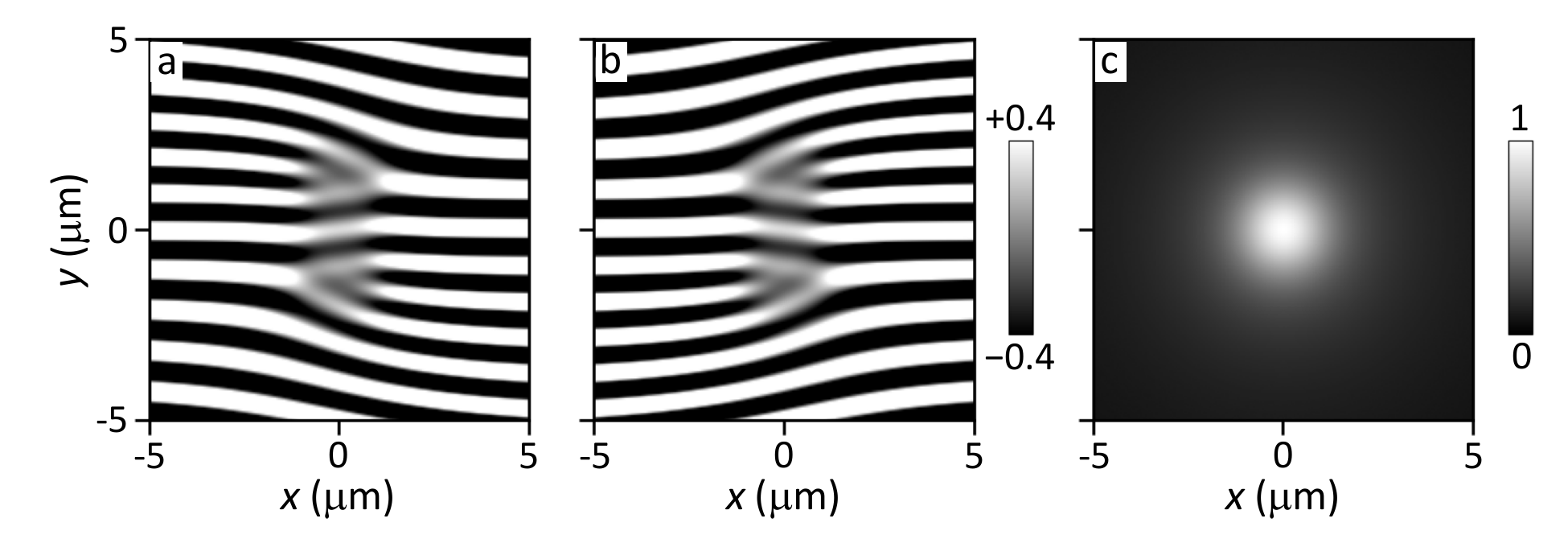}
\caption{Simulated exciton interference pattern with adjacent interference dislocations. 
(a, b) Interference pattern $I_{\rm interf}(x, y)$ for the emission spot generated by the Gaussian-shape exciton source with width $\sigma = 0.7$~$\mu$m: $P(\mathbf{r}_{\rm s}) = e^{-|\mathbf{r}_{\rm s}|^2/2\sigma^2}$. The shift $\delta x = - 1.4$~$\mu$m (a) and $\delta x = 1.4$~$\mu$m (b). 
(c) The emission spot $I_{\rm em}(x, y)$ generated by this exciton source.
The adjacent interference dislocations are located close to the emission spot.
}
\end{center}
\label{fig:spectra}
\end{figure}

\begin{figure}
\begin{center}
\includegraphics[width=8.5cm]{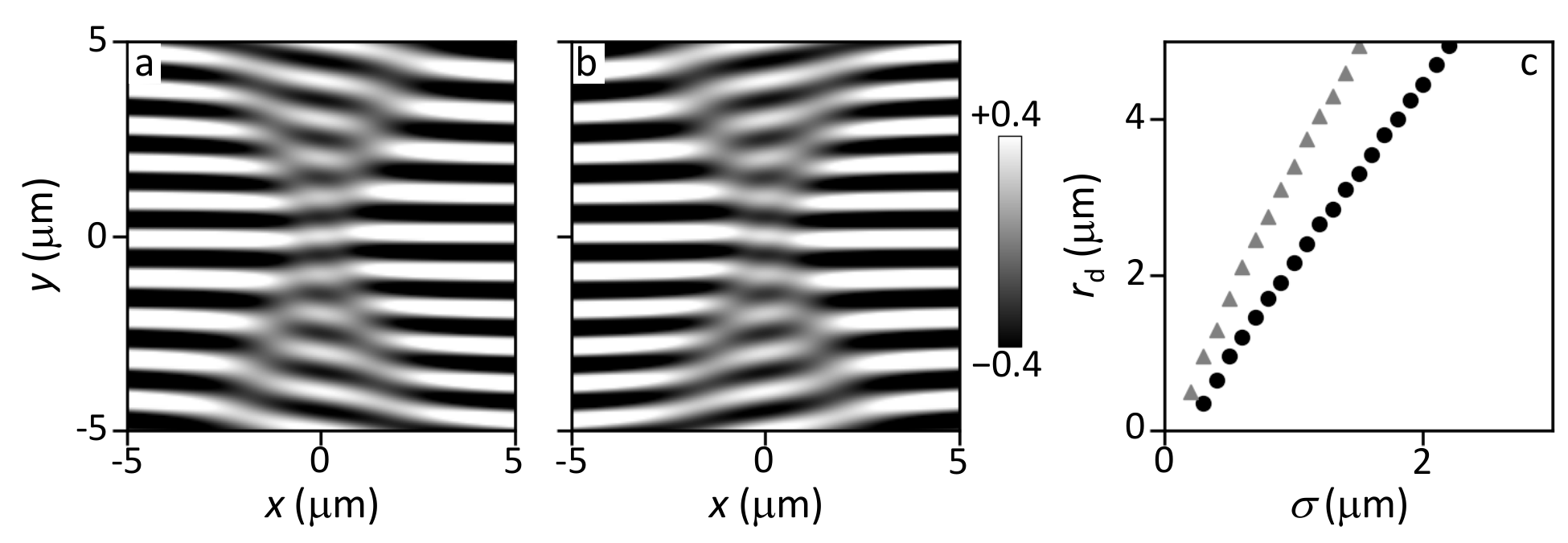}
\caption{Simulated exciton interference patterns for various emission spots. 
(a, b) Interference pattern $I_{\rm interf}(x, y)$ for the emission spot generated by the Lorentzian-shape exciton source with width $\sigma = 0.7$~$\mu$m: $P(\mathbf{r}_{\rm s}) = 1/(|\mathbf{r}_{\rm s}|^2 + \sigma^2)$. The shift $\delta x = - 1.4$~$\mu$m (a) and $\delta x = 1.4$~$\mu$m (b). 
(c) The separation of the simulated interference dislocations from the spot center vs. the spot size. The points (triangles) correspond to Gaussian-shape (Lorentzian-shape) exciton sources with width $\sigma$. 
}
\end{center}
\label{fig:spectra}
\end{figure}

The interference patterns simulated using this approach are presented in Fig.~3a,b for a Gaussian-shape exciton source. The corresponding exciton emission is shown in Fig.~3c. The Gaussian shape is close to the profile of the exciton generation spot produced by a focused laser excitation. The exciton sources of similar shapes are used in the experiments presented in Figs.~1 and 2. The simulations in Fig.~3 show the adjacent interference dislocations located close to the emission spot. These dislocations originate from the moir{\'e} effect in combined interference patterns produced by constituting parts of the emission spot described by Eqs.~1 and 2. The appearance of adjacent interference dislocations in the measured (Figs.~1, 2) and simulated (Fig.~3) interference patterns indicates that the adjacent interference dislocations in the experiment originate from the moir{\'e} effect. 

We also consider other shapes of exciton sources. Figures~4a,b show the simulated interference patterns for a Lorentzian-shape exciton source. These simulations also show adjacent interference dislocations located close to the emission spot. Figure~4c shows that the interference dislocations stay adjacent to the emission spot for various sizes of the spot. The existence of adjacent interference dislocations for various shapes (Figs.~3a,b and 4a,b) and sizes (Fig.~4c) of the emission spot indicates that the adjacent interference dislocations can appear in interference images for various emission spots. 

Multiple emission spots can produce a pattern of adjacent interference dislocations. A local enhancement of emission intensity can produce adjacent interference dislocations. In general, the adjacent interference dislocations can appear in interference images for various spatially modulated emission patterns. 

The appearance of adjacent interference dislocations does not require coherence between the parts of the emission spot: The addition of intensities from the parts of the emission spot in Eqs.~1 and 2 corresponds to the lack of coherence between the exciton waves originating from the parts of the emission spot. This shows that the adjacent interference dislocations can be observed in a classical system. This makes them different from interference dislocations observed in the condensates, such as the interference dislocations originating from quantized vortices~\cite{Inouye2001} and isolated interference dislocations originating from the moir{\'e} effect in combined interference patterns of condensate matter waves propagating long distances away from the exciton sources~\cite{Leonard2021}. 

In summary, we studied interference dislocations adjacent to an emission spot in an interference pattern. The simulations show that the adjacent interference dislocations appear due to the moir{\'e} effect in combined interference patterns produced by constituting parts of the emission spot.

\vskip5mm
{\bf Acknowledgements} 

We thank John~Schaibley, Brian~LeRoy, Jacob~Cutshall, and Misha~Fogler for discussions. The experiments were supported by the Department of Energy, Office of Basic Energy Sciences, under award DE-FG02-07ER46449. The heterostructure manufacturing, simulations, and analysis were supported by NSF Grants 1905478 and 2516006.

\end{document}

\date{\today}

\title{Supporting Information for

Interference dislocations adjacent to emission spot}

\author{J.~R.~Leonard}
\affiliation{Department of Physics, University of California San Diego, La Jolla, CA 92093, USA}
\author{L.~H.~Fowler-Gerace}
\affiliation{Department of Physics, University of California San Diego, La Jolla, CA 92093, USA}
\author{Zhiwen~Zhou}
\affiliation{Department of Physics, University of California San Diego, La Jolla, CA 92093, USA}
\author{E.~A.~Szwed}
\affiliation{Department of Physics, University of California San Diego, La Jolla, CA 92093, USA}
\author{D.~J.~Choksy}
\affiliation{Department of Physics, University of California San Diego, La Jolla, CA 92093, USA}
\author{L.~V.~Butov} 
\affiliation{Department of Physics, University of California San Diego, La Jolla, CA 92093, USA}

\maketitle

\renewcommand*{\thefigure}{S\arabic{figure}}

\subsection{Heterostructure}

The van der Walls MoSe$_2$/WSe$_2$ heterostructure was assembled using the dry-transfer peel technique~\cite{Withers2015} as described in Ref.~\cite{Fowler-Gerace2024}. The entire areas of the MoSe$_2$ and WSe$_2$ monolayers are covered by hBN layers. The thickness of the bottom hBN layer is $\sim 40$~nm, the thickness of the top hBN layer is $\sim 30$~nm. The MoSe$_2$ monolayer is on top of the WSe$_2$ monolayer. The twist angle between the MoSe$_2$ and WSe$_2$ monolayers $\delta \theta \sim 1.1^\circ$. The measured $g$-factor corresponds to H stacking in the MoSe$_2$/WSe$_2$ heterostructure~\cite{Seyler2019, Wozniak2020}. The measurements for 
spatially direct 
excitons were performed in the area of WSe$_2$ monolayer. 
The measurements for spatially indirect excitons (IXs) were performed in the area of MoSe$_2$/WSe$_2$ heterostructure where the WSe$_2$ monolayer and the MoSe$_2$ monolayer overlap.

\subsection{Optical measurements}

Excitons were generated by a cw HeNe laser with excitation energy 1.96~eV. 
For this nonresonant excitation, the regime of the long-range IX transport studied in Refs.~\cite{Fowler-Gerace2024, Zhou2024, Zhou2025} is not realized. 
The laser was focused to a spot size $\sim 2$~$\mu$m. The emission of excitons in the WSe$_2$ monolayer or the emission of IXs in the MoSe$_2$/WSe$_2$ heterostructure was spectrally selected by a long-pass filter 700~nm or 900~nm, respectively. 

The interference patterns were measured using a liquid-nitrogen-cooled CCD. The experiments were performed in a variable-temperature 4He cryostat. The presented data were measured at $T = 1.7$~K. Similar forks in the IX interference patterns were observed at higher temperatures 40, 80, and 120~K.

We used a Mach-Zehnder (MZ) interferometer (Fig.~S1) to measure the interference patterns. The path lengths of the arms of the MZ interferometer were equal. The emission images produced by each of the two arms of the MZ interferometer were shifted relative to each other along $x$ (or along $y$) in the layer plane. The interference pattern $I_{\rm interf} = (I_{12} - I_1 - I_2)/(2\sqrt{I_1 I_2})$ was calculated from the measured intensity of the emission pattern $I_1$ for arm 1 open, $I_2$ for arm 2 open, and $I_{12}$ for both arms open.

\begin{figure}
\begin{center}
\includegraphics[width=8.5cm]{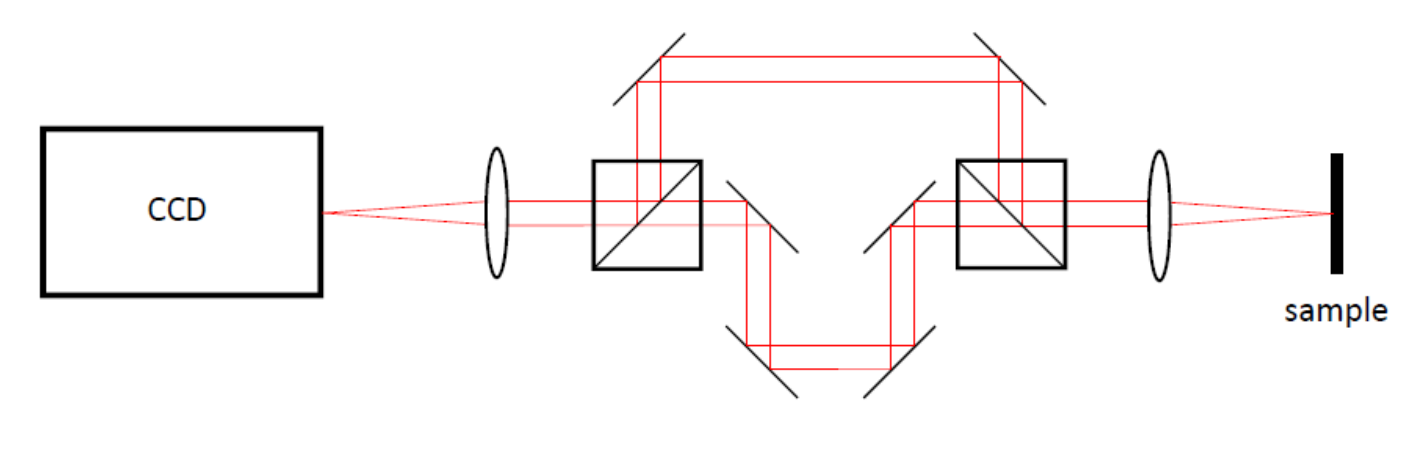}
\caption{Schematic of the shift-interferometry setup.
}
\end{center}
\label{fig:spectra}
\end{figure}

\begin{figure}
\begin{center}
\includegraphics[width=9.5cm]{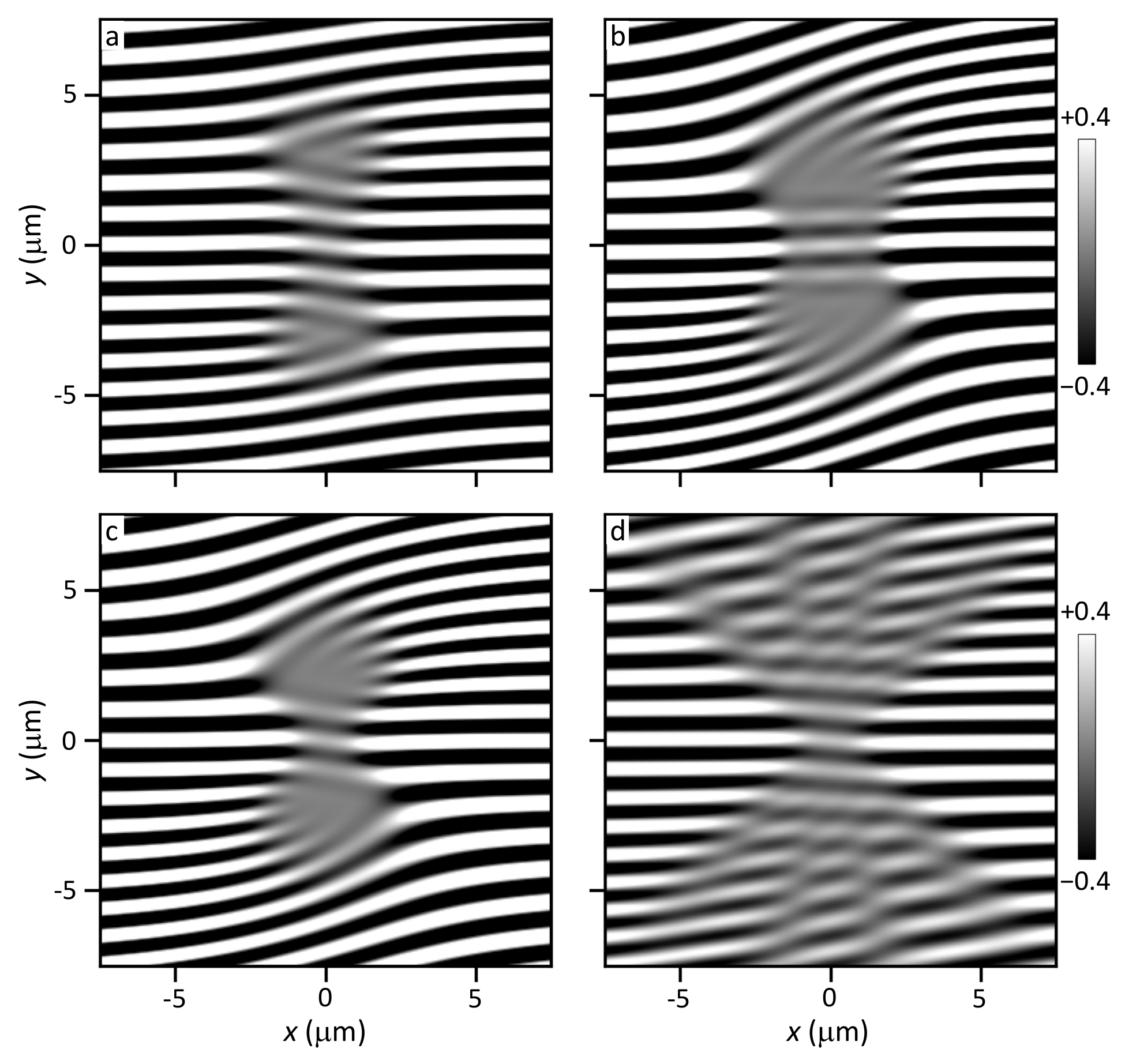}
\caption{Examples of simulated interference dislocations for $\sigma$, $\delta x$, and $k$ higher than in the main text. 
(a) The simulated interference pattern $I_{\rm interf}(x, y)$ for $\sigma = 1.4$~$\mu$m. Other parameters $k = 4.4$~$\mu$m$^{-1}$ and $\delta x = 1.4$~$\mu$m are the same as in Fig.~3 in the main text. 
(b) The simulated $I_{\rm interf}(x, y)$ for $\delta x = 3.2$~$\mu$m. Other parameters $\sigma = 0.7$~$\mu$m and $k = 4.4$~$\mu$m$^{-1}$ are the same as in Fig.~3 in the main text.  
(c) The simulated $I_{\rm interf}(x, y)$ for $k = 9$~$\mu$m$^{-1}$. Other parameters $\sigma = 0.7$~$\mu$m and $\delta x = 1.4$~$\mu$m are the same as in Fig.~3 in the main text. 
(d) The simulations in (a-c) are for Gaussian-shape excitation spot. The simulations in (d) are for Lorentzian-shape excitation spot for $k = 9$~$\mu$m$^{-1}$, $\sigma = 0.7$~$\mu$m, and $\delta x = 1.4$~$\mu$m.
The separation of the interference dislocations from the spot center is higher for higher $\sigma$, $\delta x$, and $k$ (compare Figs.~S2a, S2b, and S2c, respectively, with Fig.~3a). 
Figures~S2b--S2d show interference dislocations with 4 prongs. 
More detailed dependence of interference patterns on  $\sigma$, $\delta x$, and $k$ is shown in the movies~\cite{movie1, movie2, movie3}. 
}
\end{center}
\label{fig:spectra}
\end{figure}

\subsection{Simulated exciton interference dislocations for various parameters}

The simulations in the main text are performed for the shift $\delta x = 1.4$~$\mu$m and the 
exciton momentum $k = 4.4$~$\mu$m$^{-1}$. The interference dislocations are observed in the simulations for various $\delta x$ and $k$. 
The simulations show that the interference dislocations appear in the interference patterns with increasing $k$ and with increasing $\delta x$ and the separation of the interference dislocations from the spot center increases both with $k$ and $\delta x$, as shown in Movies~1 and 2~\cite{movie1, movie2}.  
The simulations also show that that the number of prongs in the interference dislocations increases both with $k$ and $\delta x$, as shown in Movies~1 and 2~\cite{movie1, movie2}.  
As outlined in the main text, the simulations also show that the separation of the interference dislocations from the spot center increases with the spot size $\sigma$, as shown in Movie~3~\cite{movie3}.
Examples of interference patterns for $k$, $\delta x$, and $\sigma$ higher than in the main text are shown in Fig.~S2. 

For the simulations in Figs.~3 and 4 in the main text, the excitation spot size, the period of interference fringes, and the shift are taken from the experiments and the exciton momentum $k$ is a fit parameter chosen to fit the distance from the interference dislocations to the emission spot center. The obtained $k$ also gives the dislocation prong number that agrees with the experiments: The interference patterns in the experiment (Figs.~1 and 2 in the main text) and in the simulations (Figs.~3 and 4 in the main text) show interference dislocations with two prongs.  

When the interference dislocations appear close to the excitation spot they can be called the adjacent interference dislocations and described by Eq.~1 in the main text. In this regime, the appearance of dislocations in interference patterns does not require the exciton mean free time to be long and adjacent interference dislocations can appear in classical systems with spatially modulated emission as outlined in the main text. 

In contrast, the interference dislocations at substantial distances from the exciton sources are isolated interference dislocations. The appearance of isolated dislocations in interference patterns at the locations determined by the long-range ballistic exciton transport and requiring the long exciton mean free time can evidence exciton superfluidity as outlined in Ref.~\cite{Leonard2021}. In the simulations, a crossover from adjacent interference dislocations to isolated interference dislocations can occur with increasing separation of the interference dislocation from the exciton source.

\begin{figure}
\begin{center}
\includegraphics[width=10cm]{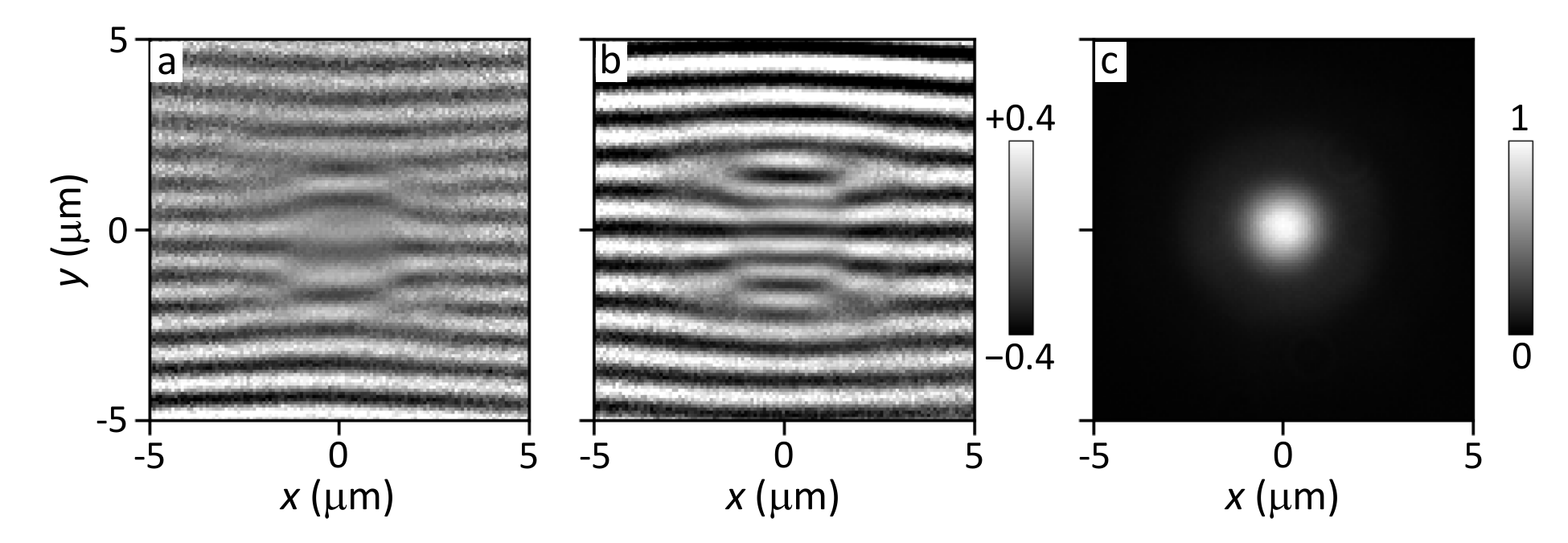}
\caption{Interference patterns for excitons in WSe$_2$ monolayer: $\delta y$-shifts. 
(a, b) Measured interference pattern $I_{\rm interf}(x, y)$ for the shift $\delta y = - 1.4$~$\mu$m (a) and $\delta y = 1.4$~$\mu$m (b). 
(c) The exciton emission spot. 
A compression (a) and expansion (b) of the pattern of interference fringes with fork-like features are observed in the area of exciton emission spot.
}
\end{center}
\label{fig:spectra}
\end{figure}

\begin{figure}
\begin{center}
\includegraphics[width=10cm]{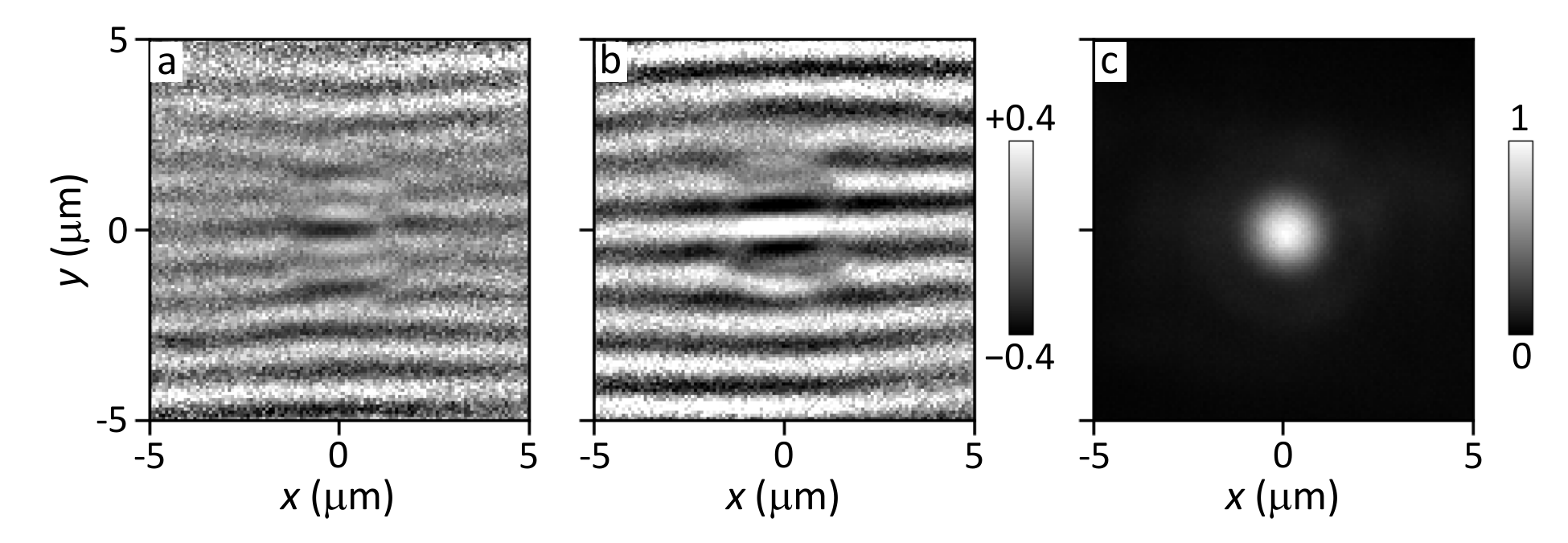}
\caption{Interference patterns for IXs in MoSe$_2$/WSe$_2$ heterostructure: $\delta y$-shifts. 
(a, b) Measured interference pattern $I_{\rm interf}(x, y)$ for the shift $\delta y = - 1.4$~$\mu$m (a) and $\delta y = 1.4$~$\mu$m (b). 
(c) The IX emission spot. 
A compression (a) and expansion (b) of the pattern of interference fringes with fork-like features are observed in the area of IX emission spot.
}
\end{center}
\label{fig:spectra}
\end{figure}

\begin{figure}
\begin{center}
\includegraphics[width=10cm]{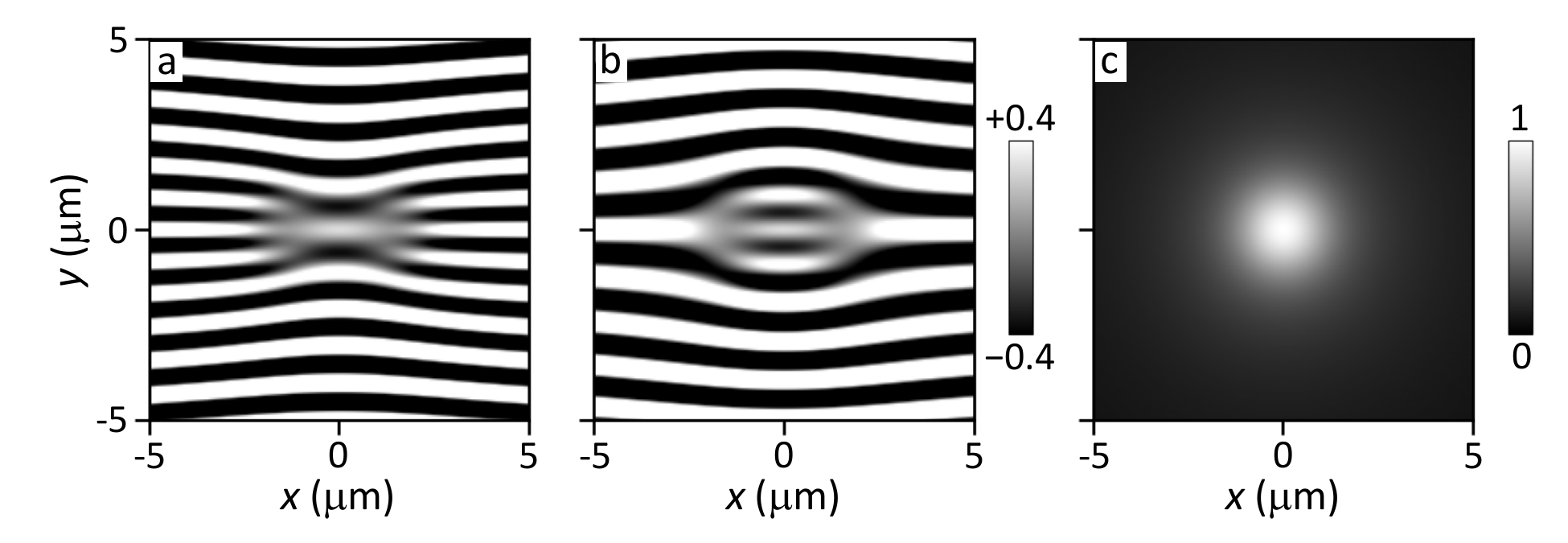}
\caption{Simulated interference patterns for excitons: $\delta y$-shifts. 
(a, b) Interference pattern $I_{\rm interf}(x, y)$ for the emission spot generated by the Gaussian-shape exciton source with width $\sigma = 0.7$~$\mu$m. The shift $\delta y = - 1.4$~$\mu$m (a) and $\delta y = 1.4$~$\mu$m (b).  $k = 4.4$~$\mu$m$^{-1}$. 
(c) The emission spot $I_{\rm em}(x, y)$ generated by this exciton source.  
The simulations show a compression (a) and expansion (b) of the pattern of interference fringes with fork-like features in the area of exciton emission spot.
}
\end{center}
\label{fig:spectra}
\end{figure}

\subsection{Interference patterns for shifts normal to interference fringes}

The experiments and simulations in Figs.~1-4, and S2 correspond to $\delta x$ shifts in the $x$ direction parallel to the interference fringes. We also measured and simulated interference patterns for $\delta y$ shifts in the $y$ direction normal to the interference fringes. Both for excitons in the WSe$_2$ monolayer (Fig.~S3) and IXs in the MoSe$_2$/WSe$_2$ heterostructure (Fig.~S4), the experiments show a compression (Figs.~S3a, S4a) and an expansion (Figs.~S3b, S4b) of the pattern of interference fringes with fork-like features in the area of emission spot. The simulations using Eq.~1 in the main text with $\delta {\bf r} = \delta {\bf y}$ show similar interference patterns (Fig.~S5).

\subsection{Shifts of interference fringes along Airy rings}

The measured interference patterns also show the features associated with the Airy rings. Figure~S6 shows the interference patterns for IXs in the MoSe$_2$/WSe$_2$ heterostructure for various $\delta x$ shifts. The black circles in Fig.~S6(g-l) indicate the dark rings in Airy patterns shifted by $\delta x$ relative to each other. The radius of the Airy rings 
is given by the numerical aperture of the lens. The interference patterns show the shifts of interference fringes along the Airy rings (Fig.~S6). 
The shifts originate from the sign change of the light field at the dark Airy ring.
These shifts complicate the interference patterns, however, the interference dislocations are clearly observed (Fig.~S6). 

\begin{figure}
\begin{center}
\includegraphics[width=14cm]{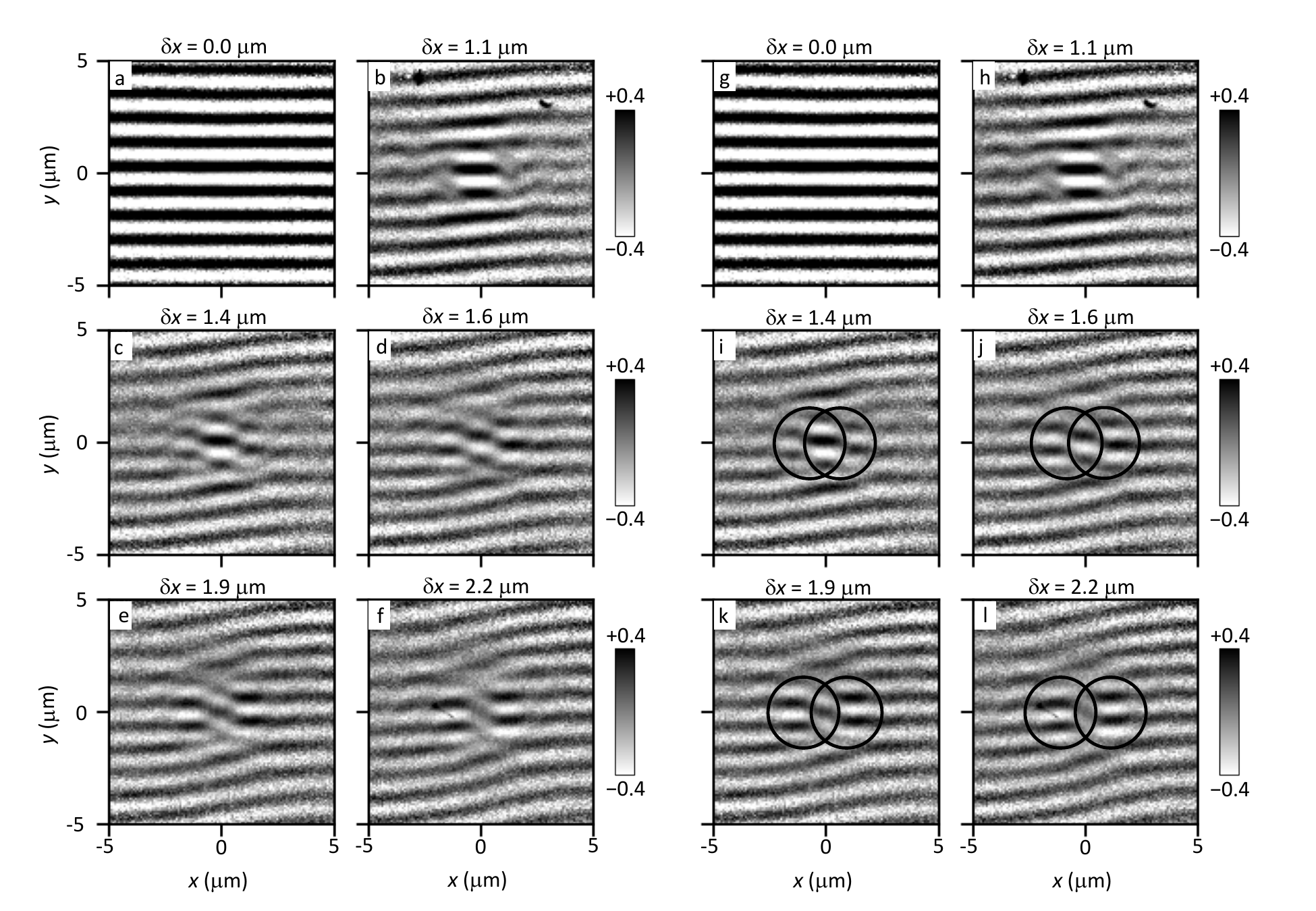}
\caption{Interference patterns for IXs in MoSe$_2$/WSe$_2$ heterostructure: $\delta x$-shift dependence. 
(a-f) Measured $I_{\rm interf}(x, y)$ vs. the shift $\delta x$. 
The IX emission spot is shown in Fig.~2c. 
(g-l) Same interference patterns as in (a-f) with the black circles indicating the Airy rings shifted by $\delta x$ relative to each other. 
The interference patterns show the shifts of interference fringes along the Airy rings.
}
\end{center}
\label{fig:spectra}
\end{figure}